\documentclass[11pt,twoside]{article}
\usepackage{fancyhdr}
\usepackage[colorlinks,citecolor=blue,urlcolor=blue,linkcolor=blue,bookmarks=false]{hyperref}
\usepackage{amsfonts,epsfig,graphicx}
\usepackage{afterpage}
\usepackage{comment}
\usepackage{amsmath,amssymb,amsthm} 
\usepackage{fullpage}
\usepackage[T1]{fontenc} 
\usepackage{epsf} 
\usepackage{graphics} 
\usepackage{amsfonts,amsmath}
\usepackage{mathtools}
\usepackage{natbib} 
\usepackage{psfrag,xspace}
\usepackage{color,etoolbox}
\usepackage{xcolor}
\usepackage{subcaption} 
\usepackage{dsfont}
\usepackage[page]{appendix}
\usepackage{thmtools}
\usepackage{soul} 

\def\ind{\perp\!\!\!\perp}

\newcommand{\Pb}{\mathbb{P}}
\newcommand{\Pn}{\mathbb{P}_n}

\newcommand{\E}{\mathbb{E}}

\def\odds{\mathrm{odds}}

\DeclareSymbolFont{bbold}{U}{bbold}{m}{n}
\DeclareSymbolFontAlphabet{\mathbbold}{bbold}
\newcommand{\one}{\mathbbold{1}}

\newtheorem{lemma}{Lemma}
\newtheorem{corollary}{Corollary}
\newtheorem{proposition}{Proposition}

\newtheorem{remark}{Remark}
\theoremstyle{definition}

\theoremstyle{remark}

\usepackage{pgf,tikz}
\usetikzlibrary{positioning}
\usetikzlibrary{shapes}
\usepackage{pgfplots}
\pgfplotsset{compat=1.18}
\usetikzlibrary{matrix}       
\usetikzlibrary{arrows,shapes,positioning,shadows,trees,arrows.meta}

\begin{document}

\def\spacingset#1{\renewcommand{\baselinestretch}%
{#1}\small\normalsize} \spacingset{1}

\raggedbottom
\allowdisplaybreaks[1]

\renewcommand\thmcontinues[1]{Continued}


\title{\vspace*{-.4in} {Discussion of ``Causal and counterfactual views of missing data models'' by Razieh Nabi, Rohit Bhattacharya, Ilya Shpitser, and James M. Robins}} \author{Alexander W. Levis \& Edward H. Kennedy \\  \\
  Department of Statistics \& Data Science, \\
  Carnegie Mellon University \\ \\
  \texttt{alevis@cmu.edu};
  \texttt{edward@stat.cmu.edu}
  \date{} }
    
  \maketitle
  \vspace{5mm}

\section{Introduction}
We congratulate \citet{nabi2022} on their impressive and insightful paper, which illustrates the benefits of using causal/counterfactual perspectives and tools in missing data problems.  This paper represents an important approach to missing-not-at-random (MNAR) problems, exploiting nonparametric independence restrictions for identification, as opposed to parametric/semiparametric models, or resorting to sensitivity analysis. Crucially, the authors represent these restrictions with missing data directed acyclic graphs (m-DAGs), which can be useful to determine identification in complex and interesting MNAR models. In this discussion we consider (i) how/whether other tools from causal inference could be useful in missing data problems, (ii) problems that combine missing data and causal inference together, and (iii) some work on estimation in one of the authors' example MNAR models. \\

\section{Other causal identification tools}

 The graphical arguments used in~\citet{nabi2022} stem from a powerful analogy to a widely used set of tools in the causal inference literature. In drawing this connection, their work raises the question: \textit{which other tools and methodologies can be borrowed from the causal literature and bear fruit for missing data problems}? \\
 
 The focus of \citet{nabi2022} is---for the most part---on characterizing identification of the full data law, $p(L^{(1)}, R)$, under sets of (factual and counterfactual) conditional independence relations implied by a directed acyclic graph and its associated structural model. However, it is often sufficient for practical purposes to identify and estimate certain functions or functionals of this distribution, say the marginal ``target'' data mean functional $\mathbb{E}(L^{(1)})$, rather than the entire full law. Moreover, although nonparametric structural models and conditional independence relations represent a large and important set of possible restrictions on the full data law, other plausible structural restrictions may partially or point identify certain estimands of interest. Indeed, both \textit{missing data instrumental variables} and \textit{shadow variables} are ideas born out of the causal inference literature---the latter analogous to a negative control outcome variable---which move beyond conditional independence restrictions and can yield partial or full identification of certain functionals of the target law.\\


 Instrumental variables (IVs), while being much more widely adopted as tools for causal inference, have been increasingly studied in recent years for use in missing data problems~\citep{tchetgen2017, sun2018b}, having been proposed for such purposes at least as early as~\citet{heckman1979}. Just as IVs can help identify causal effects in the presence of unmeasured exposure-outcome confounding, IVs for missing data can help identify aspects of the target law when the data are missing not at random (MNAR), i.e., missingness status is related to the missing variables themselves, such that adjusting for measured covariates is not sufficient. Specifically, an IV for a partially missing outcome is a completely observed cause of missingness that is unrelated to the outcome conditional on measured covariates~\citep{tchetgen2017}. In the statistical causal inference literature, IVs have been used to partially identify causal effects~\citep{balke1997, swanson2018, levis2025}. Point identification of the average treatment effect is possible under additional \textit{homogeneity} assumptions, which rule out certain kinds of effect modification by the unmeasured confounders~\citep{wang2018, hernan2020causal, levis2024roadmap}. Alternatively, under a \textit{monotonicity} assumption, the average treatment effect among compliers---subjects who would be exposed under encouragement from the IV but not otherwise---is nonparametrically identified~\citep{imbens1994}. In the missing data context, while the homogeneity approach has been employed~\citep{tchetgen2017}, the types of nonparametrically identified local subgroup effects that result from monotonicity seem to not yet have been explored. More broadly, ideas from principal stratification~\citep{frangakis2002} may be interesting to consider for missing data problems. \\

 As a parallel notion to IVs, the idea of negative control outcomes~\citep{tchetgen2014, park2024} for handling unmeasured confounding in causal inference has been recently proposed as a strategy for MNAR outcomes, under the name of ``shadow variables''~\citep{miao2015, miao2016, li2023}. A shadow variable is a completely measured predictor of the outcome that is independent of missingness status conditional on measured covariates and the potentially missing outcome. These structural requirements of course could be encoded in a DAG as in~\citet{nabi2022}, but on their own are not sufficient for identification of the full law, i.e., the algorithms in~\citet{nabi2022} would (rightfully) fail. Instead, the shadow variable conditions imply certain moment restrictions on the missingness mechanism, which guarantee identification of the means of functions of the missing variable only when brought together with ``bridge'' or ``representer'' assumptions~\citep{li2023}. Interestingly, recent advances in the proximal causal inference literature make use of a pair of negative control variables to identify causal effects under unmeasured confounding~\citep{miao2018, cui2024}. It would be interesting to consider such exploitation of multiple negative controls in the missing data context.\\

Lastly, while mentioned in Appendix S3 of \citet{nabi2022}, their work more broadly motivates the potential utility of single world intervention graphs (SWIGs; \citet{richardson2013}) in the presence of missing data. What appears particularly underexplored is the extent to which SWIGs can be used not only to elicit identification conditions for the distribution of full data variables, but also the effects of (perhaps partially missing) exposure variables. We briefly entertain this possibility in the next section.\\

\section{Simultaneous intervention on treatment and missingness}

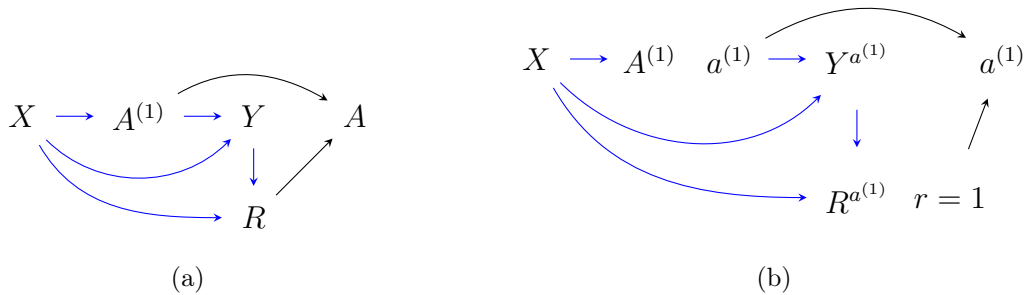
\begin{figure}[ht]
\centering
\begin{subfigure}{.5\textwidth}
\begin{center}
\large{\begin{tikzpicture}[%
        ->,
        >=stealth,
        node distance=0.5cm,
        pil/.style={
          ->,
          thick,
          shorten =2pt,},
        regnode/.style={circle, draw=none, fill=none, thick, minimum size=8mm},
        bluenode/.style={circle, draw=blue, fill=none, thick, minimum size=8mm},
        boxednode/.style={rectangle, draw=black, fill=none, thick, minimum size=8mm},
        leftsplitnode/.style={semicircle, draw=black, fill=none, thick, minimum size=9mm, shape border rotate=90},
        rightsplitnode/.style={semicircle, draw=red, fill=none, thick, minimum size=9mm, shape border rotate=270},
        rednode/.style={circle, draw=red, fill=none, thick, minimum size=8mm},
        redbox/.style={rectangle, draw=red, fill=none, thick, minimum size=8mm},
        ]
        \node[regnode] (1) {$X$};
        \node[regnode, right=of 1] (2) {$A^{(1)}$};
        \node[regnode, right=of 2] (3) {$Y$};
        \node[regnode, below=of 3] (4) {$R$};
        \node[regnode, right=of 3] (5) {$A$};
        \draw [blue, ->] (1) to (2);
        \draw [blue, ->] (2) to (3);
        \draw [blue, ->] (3) to (4);
        \draw [blue, ->] (1) to [out=300, in=180] (4);
        \draw [blue, ->] (1) to [out=315, in=225] (3);
        \draw [->] (4) to (5);
        \draw [->] (2) to [out=30, in=150] (5);
\end{tikzpicture}}
\caption{}
\label{fig:miss-exp-DAG}
\end{center}
\end{subfigure}%
\begin{subfigure}{.5\textwidth}
\begin{center}
\large{\begin{tikzpicture}[%
        ->,
        >=stealth,
        node distance=0.5cm,
        pil/.style={
          ->,
          thick,
          shorten =2pt,},
        regnode/.style={circle, draw=none, fill=none, thick, minimum size=8mm},
        bluenode/.style={circle, draw=blue, fill=none, thick, minimum size=8mm},
        boxednode/.style={rectangle, draw=black, fill=none, thick, minimum size=8mm},
        leftsplitnode/.style={semicircle, draw=none, fill=none, thick, minimum size=9mm, shape border rotate=90},
        rightsplitnode/.style={semicircle, draw=none, fill=none, thick, minimum size=9mm, shape border rotate=270},
        rednode/.style={circle, draw=red, fill=none, thick, minimum size=8mm},
        redbox/.style={rectangle, draw=red, fill=none, thick, minimum size=8mm},
        ]
        \node[regnode] (1) {$X$};
        \node[leftsplitnode, right=of 1] (2) {$A^{(1)}$};
        \node[rightsplitnode, right=1mm of 2] (3) {$a^{(1)}$};
        \node[regnode, right=of 3] (4) {$Y^{a^{(1)}}$};
        \node[regnode, below=of 4] (5) {$R^{a^{(1)}}$};
        \node[regnode, right=-1.2mm of 5] (6) {$r = 1$};
        \node[regnode, right=7mm of 4] (7) {$a^{(1)}$};
        \draw [blue, ->] (1) to (2);
        \draw [blue, ->] (3) to (4);
        \draw [blue, ->] (4) to (5);
        \draw [blue, ->] (1) to [out=300, in=180] (5);
        \draw [blue, ->] (1) to [out=315, in=225] (4);
        \draw [->] (6) to (7);
        \draw [->] (3) to [out=30, in=150] (7);
\end{tikzpicture}}
\caption{}
\label{fig:miss-exp-SWIG}
\end{center}
\end{subfigure}%
\caption{m-DAG and m-SWIG for missing point exposure}
\end{figure}

To confine our discussion within the limits of a concrete example, consider the observational point exposure setting with partially missing exposures. Namely, suppose we observe $O = (X, R, RA, Y)$, where $X$ are a set of baseline confounders, $A$ is a partially missing exposure variable with $R = \mathds{1}(A \text{ observed})$, and $Y$ is an outcome of interest. This setting was considered in \citet{williamson2012} and \citet{kennedy2020a}. Adopting the notational conventions of~\citet{nabi2022}, we can equivalently posit the existence of $A^{(1)}$, the exposure variable under an intervention that sets $R = 1$, and write $O = (X, R, A, Y)$ where $A = RA^{(1)} + (1 - R)\text{``$?$''}$. Suppose interest lies in identifying and estimating $\mathbb{E}(Y^{a^{(1)}})$, i.e., the mean potential outcome under intervention that sets $A^{(1)}$ to $a^{(1)}$. Moreover, suppose that we are in the missing at random (MAR) setting where 
\begin{equation}\label{eq:MAR}
    R \ind A^{(1)} \mid X, Y,
\end{equation}
a particular example for which is illustrated in Figure~\ref{fig:miss-exp-DAG}, adopting the graphical conventions used in~\citet{nabi2022}. \\

\bigskip

Under either the nonparametric structural equation model with independent errors~\citep{pearl2009} or the finest fully randomized causally interpretable structured tree graph~\citep{robins2010}, the m-DAG in Figure~\ref{fig:miss-exp-DAG} implies 
\begin{equation}\label{eq:NUC}
    A^{(1)} \ind Y^{a^{(1)}} \mid X.
\end{equation} Combining this no unmeasured confounding (NUC) condition with the MAR condition~\eqref{eq:MAR}, \citet{kennedy2020a} shows via Bayes' rule that
\begin{equation}\label{eq:ident}
    \mathbb{E}(Y^{a^{(1)}} \mid X) = \frac{\mathbb{E}(Y\lambda_{a^{(1)}}(X, Y) \mid X)}{\mathbb{E}(\lambda_{a^{(1)}}(X, Y) \mid X)},
\end{equation}
where $\lambda_a(X, Y) = \mathbb{P}[A = a \mid X, Y, R = 1]$. \\

Note that whereas~\eqref{eq:MAR} can be read off from the m-DAG immediately via d-separation, the NUC condition~\eqref{eq:NUC} cannot. In fact, this is a key innovation of SWIGs, in that under the above-mentioned causal models, this conditional independence on counterfactuals can be determined via d-separation after a node-splitting operation \citep{richardson2013}. Consider the intervention that sets treatment to $a^{(1)}$ and the missingness indicator to 1. In this case, to draw the corresponding SWIG, the node $A^{(1)}$ is split into random ($A^{(1)}$) and fixed ($a^{(1)}$) components, with the former retaining all incoming arrows from the original DAG, and the latter retaining all outgoing arrows; any (non-deterministic) descendant of the fixed node gets the counterfactual label $a^{(1)}$. The SWIG (or, perhaps, ``m-SWIG'') corresponding to this intervention is displayed in Figure~\ref{fig:miss-exp-SWIG}, where we performed the same operation for the missingness indicator. \\

A couple of observations are worth noting. First, we can apparently in this example say more than~\eqref{eq:NUC}, namely that $A^{(1)} \ind (Y^{a^{(1)}}, R^{a^{(1)}}) \mid X$, since $X$ is the only parent of $A^{(1)}$ in the SWIG. Interestingly, although we are in a MAR setting, the missingness indicator $R$ is a descendant of $A^{(1)}$ (through the path $A^{(1)} \to Y \to R$) and thus the counterfactual missingness indicator $R^{a^{(1)}}$ appears on the SWIG. Second, by construction, m-DAGs and corresponding m-SWIGs contain deterministic relationships that follow from consistency, e.g., the variable $A$ is a non-random function of $A^{(1)}$ and $R$. Thus, we could follow~\citet{nabi2022} and collapse $(A^{(1)}, A)$ as a single variable when the missingness indicator is equal to 1, though in Figure~\ref{fig:miss-exp-SWIG} we keep the deterministic $A^{A^{(1)}=a^{(1)}, r = 1} \equiv a^{(1)}$ for clarity. Such deterministic relationships can subtly complicate the characterization of target law identification. For instance, the ``twin networks'' of \citet{balke1994} and \citet{pearl2009} were an earlier graphical attempt to represent counterfactual independence relations, but for which d-separation failed to be complete due to the presence of deterministic relationships~\citep{richardson2013, shpitser2022}. It is plausible that such complications will not arise under the convention adopted by the authors whereby $L^{(1)}$ cannot be descendants of $R$ and $L$. More generally, whether non-random variable dependencies have an impact on d-separation statements in m-SWIGs may be an important avenue for future inquiry. \\

While the above example may not be particularly interesting from an identification perspective (since~\eqref{eq:ident} can be obtained by elementary means), it raises the question of whether SWIGs can provide insight into more complicated problems at the intersection of causal inference and missing data---say with partially missing confounders, treatments, and outcomes---or even problems purely in the missing data realm. The simplicity of SWIGs has made them increasingly popular in applied causal settings, and we hope to see them more fully investigated for missing data. \\

\section{Statistical estimation considerations}

Moving away from the usual missing at random assumptions can yield non-standard identification formulas for target parameters. As a result, efficient nonparametric estimation in these alternative paradigms often remains unexplored. We briefly illustrate some interesting challenges by considering estimation of mean outcomes in the permutation model first proposed by \citet{robins1997non} and used by \citet{nabi2022} as a primary  example. \\

We relabel $Y \equiv L_1$ and $X \equiv L_2$, so that in the authors' motivating example,  $Y^{(1)}$ is the (possibly unmeasured) HIV status and $X^{(1)}$ denotes HIV risks and fears. The missing data DAG for this model is reproduced in Figure \ref{fig:permdag}  below. \\

\begin{figure}[ht]
\centering
\begin{tikzpicture}[
  node distance=1.3cm and 2cm,
  every node/.style={font=\itshape},
  bluearrow/.style={->, thick, blue, >=Stealth},
  grayarrow/.style={->, thick, gray, >=Stealth}
]

\node (Y1) at (0,4) {$Y^{(1)}$};
\node (R1)  at (0,2.3) {$R_1$};
\node (L1)  at (0,0.5) {$Y$};

\node (X1) at (3.5,4) {$X^{(1)}$};
\node (R2)  at (3.5,2.3) {$R_2$};
\node (L2)  at (3.5,0.5) {$X$};

\draw[bluearrow] (Y1) -- (X1);
\draw[bluearrow] (X1) -- (R1);
\draw[bluearrow] (R1) -- (R2);
\draw[bluearrow] (L1) -- (R2);

\draw[grayarrow] (Y1) to[out=-120, in=120] (L1);
\draw[grayarrow] (X1) to[out=-60, in=60] (L2);
\draw[grayarrow] (R1) -- (L1);
\draw[grayarrow] (R2) -- (L2);
\end{tikzpicture}
\caption{Directed acyclic graph for the permutation missingness model.}
\label{fig:permdag}
\end{figure}
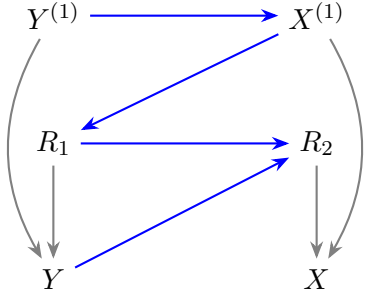

\bigskip

Under this model we have $R_1 \ind Y^{(1)} \mid X^{(1)}$ and $R_2 \ind (Y^{(1)}, X^{(1)}) \mid Y, R_1$; \citet{nabi2022} show how this implies that the joint distribution of $X^{(1)}$ and $Y^{(1)}$ is identified as
{\small
\begin{align*}
    & p_{X^{(1)}, Y^{(1)}}(x,y) \\
    &= \frac{p(x,y, R_1 = 1, R_2 = 1) \left\{p(x \mid R_1 = 0, R_2 = 1) \mathbb{P}[R_1 = 0] + \int_{\mathcal{Y}} p(x \mid R_1 = 1, Y = y', R_2 = 1)p(R_1 = 1, y') \, dy'\right\}}{\mathbb{P}[R_2 = 1 \mid R_1 = 1, Y = y] \int_{\mathcal{Y}} p(x \mid R_1 = 1, Y = y', R_2 = 1)p(R_1 = 1, y') \, dy'} \\
    &= p(R_1 = 1, y)p(x \mid R_1 = 1, Y = y, R_2 = 1) \left\{1 + \frac{p(x \mid R_1 = 0, R_2 = 1) \mathbb{P}[R_1 = 0]}{\int_{\mathcal{Y}} p(x \mid R_1 = 1, Y = y', R_2 = 1)p(R_1 = 1, y') \, dy'}\right\}.
\end{align*}
}%
Suppose we do not wish to estimate the entire joint distribution, but instead $\psi = \mathbb{E}(Y^{(1)})$, e.g., the proportion of HIV-positive subjects in the previous example. The next proposition gives the identifying expression for this parameter under the permutation model. \\

\begin{proposition}
\label{prop:psi}
Assume the permutation missing model in Figure \ref{fig:permdag}. Then $\psi=\E(Y^{(1)})$ is given by
$$\psi = \mathbb{P}(R_1 = 1)\mathbb{E}(Y \mid R_1 = 1) + \mathbb{P}(R_1 = 0) \theta ,$$ where
\[\theta = \E(Y^{(1)} \mid R_1=0) = \mathbb{E}\left(\frac{\mathbb{E}\left\{Y / \zeta(Y) \mid R_1 = 1, R_2 = 1, X\right\}}{\mathbb{E}\left\{1 / \zeta(Y) \mid R_1 = 1, R_2 = 1, X\right\}}\, \middle| \, R_1 = 0, R_2 = 1\right)\]
for $\zeta(Y) = \mathbb{P}[R_2 = 1 \mid R_1 = 1, Y]$.
\end{proposition}

\bigskip

\begin{proof}
By identification of the full law above, we have
{\small
\begin{align*}
    \psi &= \int_{\mathcal{Y}} y \int_{\mathcal{X}} p_{X^{(1)}, Y^{(1)}}(x,y) \, dx \, dy \\
    &= \int_{\mathcal{Y}} y\, p(R_1 = 1, y)\, dy \\
    & \quad \quad + \mathbb{P}[R_1 = 0]\int_{\mathcal{X}} p(x \mid R_1 = 0, R_2 = 1) \frac{\int_{\mathcal{Y}} y \, p(y \mid R_1 = 1)p(x \mid R_1 = 1, Y = y, R_2 = 1) \, dy}{\int_{\mathcal{Y}} \, p(y \mid R_1 = 1)p(x \mid R_1 = 1, Y = y, R_2 = 1) \, dy}\, dx\\
    &= \mathbb{P}[R_1 = 1] \cdot \mathbb{E}(Y \mid R_1 = 1) \\
    & \quad \quad + \mathbb{P}[R_1 = 0] \cdot \mathbb{E}\left(\frac{\int_{\mathcal{Y}} y \, p(y \mid R_1 = 1)p(X \mid R_1 = 1, Y = y, R_2 = 1) \, dy}{\int_{\mathcal{Y}} \, p(y \mid R_1 = 1)p(X \mid R_1 = 1, Y = y, R_2 = 1) \, dy} \, \middle| \, R_1 = 0, R_2 = 1\right)
\end{align*}
}%
Now by Bayes' rule $p(X \mid R_1 = 1, Y = y, R_2 = 1) = \frac{p(y \mid R_1 = 1, R_2 = 1, X)p(X\mid R_1 = 1, R_2 = 1)}{p(y \mid R_1 = 1, R_2 = 1)}$ and $\frac{p(y \mid R_1 = 1)}{p(y \mid R_1 = 1, R_2 = 1)} = \frac{\mathbb{P}[R_2 = 1 \mid R_1 = 1]}{\mathbb{P}[R_2 = 1 \mid R_1 = 1, Y]}$, which gives the result. 
\end{proof}

\bigskip

\begin{remark}
We use a particular parametrization that ensures all nuisances can be represented by conditional expectations (at the expense of preserving variation independence); alternative parametrizations are of course also possible (and later we consider a different parametrization when $Y$ is binary).  \\
\end{remark}

Next we establish the nonparametric influence function of the functional $\theta$, the variance of which represents a local asymptotic minimax lower bound for estimation of this quantity. \\

\begin{lemma}
\label{lem:eif}
Let $\zeta(Y) = \mathbb{P}[R_2 = 1 \mid R_1 = 1, Y]$ as before and define
\begin{align*}
\alpha(X) &= \mathbb{E}\left(1 / \zeta(Y) \mid R_1 = 1, R_2 = 1, X\right) \\
\beta(X) &= \mathbb{E}\left(Y / \zeta(Y) \mid R_1 = 1, R_2 = 1, X\right) \\
\gamma(X) &= \mathbb{P}[R_1 = 1 \mid R_2 = 1, X] \\
\delta(Y) &= \mathbb{E}\left(\frac{1}{\alpha(X)}\frac{1 - \gamma(X)}{\gamma(X)}\left\{Y - \frac{\beta(X)}{\alpha(X)}\right\}\,\middle| \,R_1 =1, R_2 = 1, Y\right). 
\end{align*}
    Then the functional $\theta=\mathbb{E}(\beta(X) / \alpha(X) \mid R_1 = 0, R_2 = 1)$ satisfies the von Mises expansion
    \[
        \theta(\overline{\mathbb{P}}) - \theta(\mathbb{P}) = \int \varphi(o; \overline{\mathbb{P}}) \, d(\overline{\mathbb{P}} - \mathbb{P})(o) + R_{\theta}(\overline{\mathbb{P}}; \mathbb{P}),
    \]
    with influence function
    \begin{align*}
        \varphi(O;\mathbb{P}) &= \frac{1}{\mathbb{P}[R_1 = 0, R_2 = 1]}\bigg[(1 - R_1)R_2\left\{\frac{\beta(X)}{\alpha(X)} - \theta\right\} + \frac{R_1}{\alpha(X)}\frac{R_2}{\zeta(Y)}\frac{1 - \gamma(X)}{\gamma(X)}\left\{Y - \frac{\beta(X)}{\alpha(X)}\right\} \\
        & \hspace{1.7in} - R_1\left(\frac{R_2}{\zeta(Y)} - 1\right)\delta(Y)
        \bigg],
    \end{align*}
    and remainder
    \begin{align*}
        & R_{\theta}(\overline{\mathbb{P}}; \mathbb{P}) = \left\{1 - \frac{\mathbb{P}[R_1 = 0, R_2 = 1]}{\overline{\mathbb{P}}[R_1 = 0, R_2 = 1]}\right\}\left\{\theta(\overline{\mathbb{P}}) - \theta(\mathbb{P})\right\} \\
        & \quad \quad + \frac{1}{\overline{\mathbb{P}}[R_1 = 0, R_2 = 1]}\mathbb{E}_{\mathbb{P}}\bigg(R_2 \left\{\frac{\overline{\beta}(X)}{\overline{\alpha}(X)} - \frac{\beta(X)}{\alpha(X)}\right\}\left\{[1 - \gamma(X)] - \frac{\alpha(X)}{\overline{\alpha}(X)} \frac{\gamma(X)}{\overline{\gamma}(X)}[1 - \overline{\gamma}(X)]\right\} \\
        & \quad \quad \quad \quad - \frac{R_1}{\overline{\alpha}(X)}R_2 \frac{1 - \overline{\gamma}(X)}{\overline{\gamma}(X)}\left\{\frac{\overline{\beta}(X)}{\overline{\alpha}(X)} - \frac{\beta(X)}{\alpha(X)}\right\}\left\{\frac{1}{\overline{\zeta}(Y)} - \frac{1}{\zeta(Y)}\right\}\\
        & \quad \quad \quad \quad - R_1 R_2 \left\{\frac{1}{\overline{\zeta}(Y)} - \frac{1}{\zeta(Y)}\right\} \left\{\overline{\delta}(Y) - \delta(Y)\right\}\\
        & \quad \quad \quad \quad + R_1 R_2 \left(Y - \frac{\beta(X)}{\alpha(X)}\right)\left\{\frac{1}{\overline{\zeta}(Y)} - \frac{1}{\zeta(Y)}\right\} \left\{\frac{1}{\overline{\alpha}(X)}\frac{1 - \overline{\gamma}(X)}{\overline{\gamma}(X)} - \frac{1}{\alpha(X)} \frac{1 - \gamma(X)}{\gamma(X)}\right\}
        \bigg).
    \end{align*}
\end{lemma}

\bigskip

\begin{proof}
We omit details, but the result follows from calculations similar to those discussed for example in Section 4 of \citet{kennedy2022review}.
\end{proof}

\bigskip

The results above are given for arbitrary real-valued $Y$. When $Y \in \{0,1\}$ is binary, e.g., HIV status in the motivating example, then many of the expressions simplify substantially. The following results illustrate this. \\

\begin{corollary}
\label{cor:binid}
Define
\begin{align*}
\rho &= \frac{\odds(Y=1 \mid R_1=R_2=1)}{\odds(Y=1 \mid R_1=1)} \\
\xi(x) &= \odds(Y=1 \mid X=x, R_1=R_2=1) .
\end{align*}
When $Y$ is binary then the quantity $\theta$ from Proposition \ref{prop:psi} is given by
$$ \theta = \E\left\{ \frac{\xi(X)}{\rho + \xi(X)} \Bigm| R_1=0, R_2=1 \right\} . $$ 
\end{corollary}

\bigskip

The above result aids interpretation of the functional $\theta$ in the binary case. Specifically, first note that the conditional odds in the expression can also be written as a product of a density ratio and marginal odds, i.e.,  
\begin{align*}
\xi(X) &=  \odds(Y=1 \mid X, R_1=R_2=1) \\
&= \frac{d\Pb(X \mid R_1=R_2=1,Y=1)}{d\Pb(X \mid R_1=R_2=1, Y=0)} \times \odds(Y=1 \mid R_1=R_2=1) . 
\end{align*}
Therefore denoting the density ratio by 
\begin{align*}
\lambda(x) &= \frac{d\Pb(x \mid R_1=R_2=1,Y=1)}{d\Pb(x \mid R_1=R_2=1, Y=0)} 
\end{align*}
we can also write
$$ \theta = \E\left\{ \frac{\odds(Y=1 \mid R_1=1)\lambda(X)}{1 + \odds(Y=1 \mid R_1=1)\lambda(X)}  \Bigm| R_1=0, R_2=1 \right\} . $$
Now note that the term 
$$ \odds(Y=1 \mid R_1=1) \lambda(X) $$
can be interpreted via Bayes' theorem as a posterior odds, where the prior odds, $\odds(Y=1 \mid R_1=1)$, comes from the $R_1=1$ group (those initially tested in the HIV example), while the likelihood ratio $\lambda(x)$ comes from the $R_1=R_2=1$ group (those tested and for whom data on risks/fears were measured). Then the quantity
$$ \frac{\odds(Y=1 \mid R_1=1)\lambda(X)}{1 + \odds(Y=1 \mid R_1=1)\lambda(X)} $$
is just this posterior odds converted to the probability scale. Therefore this quantity reflects the posterior probability of a positive HIV test, combining the prior information from the initial testing, with the likelihood ratio from the second round of data collection on risks/fears. Finally  the outer expectation standardizes to the $R_1=0,R_2=1$ group (those not tested but with measured risks/fears). \\

The next result gives the influence function for $\theta$ in the binary case, under a different parametrization from the influence function result above for the general $Y$ case. For simplicity, and to focus ideas, we also consider the setting where the odds ratio
$\rho$ is known. We discuss how the influence function and corresponding estimators are still useful when the odds ratio is unknown after the result. \\

\begin{corollary} \label{cor:bineff}
Define the odds ratio $\rho$ and conditional odds $\xi(x)$ as in Corollary \ref{cor:binid}, and let
$$ \varpi(x) = \frac{d\Pb(x \mid R_1=0,R_2=1)}{d\Pb(x \mid R_1=R_2=1)}. $$
When $Y$ is binary and $\rho$ is known, the functional $\theta$ from Proposition \ref{prop:psi} satisfies the von Mises expansion
$$ \theta(\overline{\mathbb{P}}) - \theta(\mathbb{P}) = \int \varphi(o; \overline{\mathbb{P}}) \, d(\overline{\mathbb{P}} - \mathbb{P})(o) + R_{\theta}(\overline{\mathbb{P}}; \mathbb{P}), $$
with influence function $\varphi(O;\Pb)$ equal to
\begin{align*} 
\rho \left\{ \frac{1+\xi(X)}{\rho+\xi(X)} \right\}^2 \frac{R_1 R_2 \varpi(X)}{\Pb[R_1=R_2=1]}  \left\{ Y - \frac{\xi(X)}{1+\xi(X)} \right\}  + \frac{(1-R_1)R_2}{\Pb[R_1=0,R_2=1]} \left\{ \frac{\xi(X)}{\rho + \xi(X)}  - \theta \right\}
\end{align*}
and remainder $R_{\theta}(\overline{\mathbb{P}}; \mathbb{P})$ given by 
\begin{align*} 
& \frac{\E(R_1R_2)}{\overline\E(R_1R_2)} \int  \left( \frac{\xi}{\rho + \xi} - \frac{\overline\xi}{\rho + \overline\xi}   \right) \Big( \overline\varpi - \varpi \Big) \ dP_{11}  + \frac{\E(R_1R_2)}{\overline\E(R_1R_2)}  \int \left(  S_{2g}  +\frac{ S_{2f} \rho}{(\rho + \overline\xi)^2}   \right)\overline\varpi \ dP_{11}  \\
& \hspace{.5in} + \left\{ \frac{\E((1-R_1)R_2)}{\overline\E((1-R_1)R_2)} -  \frac{\E(R_1R_2)}{\overline\E(R_1R_2)} \right\}  \int \left(  \frac{\overline\xi}{\rho+\overline\xi} -   \frac{\xi}{\rho+\xi}  \right) \ dP_{01} \\
& \hspace{.5in} + \left\{ 1-  \frac{\E((1-R_1)R_2)}{\overline\E((1-R_1)R_2)} \right\} \Big( \overline\theta - \theta \Big) ,
\end{align*}
for $(S_{2f},S_{2g})$ second-order terms defined in the appendix, and $dP_{st}=d\Pb(x \mid R_1=s,R_2=t)$.
\end{corollary}

\bigskip

\begin{proof}
The influence function can be obtained using the facts that (i) for $g(\xi)=\frac{\xi}{\rho+\xi}$ we have $g'(\xi) = \frac{\rho}{(\rho+\xi)^2}$, and (ii) the influence function for $\xi(x)$ when $X$ is discrete is given by
\begin{align*}
 \frac{\{ 1 + \xi(X)\}^2 \one(X=x) R_1 R_2}{d\Pb(x \mid R_1=R_2=1)\Pb[R_1=R_2=1]} \big\{ Y - \Pb[Y=1 \mid X, R_1=R_2=1] \Big\} ,
\end{align*}
together with calculations similar to those discussed in Section 4 of \citet{kennedy2022review}. A derivation of the remainder expression is given in the appendix.
\end{proof}

\bigskip

The above motivates the one-step/double-machine-learning estimator
$$ \widehat\theta = \Pn \left\{ \left( \frac{1+\widehat\xi}{\rho+\widehat\xi} \right)^2 \frac{(R_1 R_2)  \rho \widehat\varpi}{\Pn(R_1=R_2=1)}  \left( Y - \frac{\widehat\xi}{1+\widehat\xi} \right)  + \frac{(1-R_1)R_2}{\Pn(R_1=0,R_2=1)} \left( \frac{\widehat\xi}{\rho + \widehat\xi} \right) \right\} . $$
When the odds ratio $\rho$ is unknown, one can simply use the above estimator with the plug-in estimator $\widehat\rho$ replacing $\rho$. Since $\rho$ can be estimated with simple sample averages, the plug-in is $\sqrt{n}$-consistent and asymptotically normal, and the resulting estimator of $\theta$ will just have an extra asymptotically linear term.

\clearpage

\section*{References}
\vspace{-1cm}
\bibliographystyle{asa}
\bibliography{bibliography.bib}

\clearpage

\begin{appendices}


\section*{Proof of Corollary \ref{cor:bineff}}

Letting $\xi = \mu/(1-\mu)$, where $\mu(x) = \mathbb{P}[Y = 1 \mid X = x, R_1 = R_2 = 1]$, denoting $dP_{r_1 r_2}(x) = d\Pb(x \mid R_1=r_1,R_2=r)$, and omitting function arguments for simplicity, note that
\begin{align*}
\int \varphi(\overline\Pb) \ d\Pb &= \int \left\{ \frac{\rho}{(\rho + \overline\xi)^2}  \Big( 1 + \overline\xi\Big)^2 \frac{R_1 R_2\overline\varpi}{\overline\E(R_1R_2)} \Big( Y - \overline\mu \Big)  + \frac{(1-R_1)R_2}{\overline\E((1-R_1)R_2)}  \left( \frac{\overline\xi}{\rho+\overline\xi} - \overline\theta \right) \right\} \ d\Pb \\
&= \frac{\E(R_1R_2)}{\overline\E(R_1R_2)}  \int \frac{\rho}{(\rho + \overline\xi)^2}  \Big( 1 + \overline\xi\Big)^2 \Big( \mu - \overline\mu \Big) \overline\varpi \ dP_{11} + \frac{\E((1-R_1)R_2)}{\overline\E((1-R_1)R_2)} \int \left(  \frac{\overline\xi}{\rho+\overline\xi} -  \overline\theta  \right) \ dP_{01} .
\end{align*}
Therefore rearranging shows that the remainder $R_{\theta}(\overline{\mathbb{P}}; \mathbb{P})$ equals
\begin{align}
 &  \frac{\E(R_1R_2)}{\overline\E(R_1R_2)}  \int \frac{\rho}{(\rho + \overline\xi)^2}  \Big( 1 + \overline\xi\Big)^2 \Big( \mu - \overline\mu \Big) \overline\varpi \ dP_{11}  + \frac{\E((1-R_1)R_2)}{\overline\E((1-R_1)R_2)} \int \left(  \frac{\overline\xi}{\rho+\overline\xi} -  \theta  \right) \ dP_{01}   \label{eq:term1rem} \\
& \hspace{.5in} + \left\{ 1-  \frac{\E((1-R_1)R_2)}{\overline\E((1-R_1)R_2)} \right\} \Big( \overline\theta - \theta \Big) . \label{eq:term2rem}
\end{align}
The term \eqref{eq:term2rem} above is second-order. Therefore consider  term \eqref{eq:term1rem}.

By the mean value theorem we have, for some $\overline\xi'$ between $\xi$ and $\overline\xi$, that 
\begin{align*}
(\mu - \overline\mu) (1+\overline\xi)^2 &= (\mu - \overline\mu) (1+\overline\xi')^2 + (\mu - \overline\mu) \Big\{ (1+\overline\xi)^2  - (1+\overline\xi')^2  \Big\} \\
&= (\xi - \overline\xi) + (\mu - \overline\mu) \Big\{ (1+\overline\xi)^2  - (1+\overline\xi')^2  \Big\}
\end{align*}
since for $\xi=f(\mu)=\frac{\mu}{1-\mu}$ we have $f'(\mu) = 1/(1-\mu)^2 = (1+\xi)^2$. Call the second term in the last line above $S_{2f}(x)$.
Similarly we have for some other $\overline\xi''$ between $\xi$ and $\overline\xi$, that
\begin{align*}
(\xi - \overline\xi) \frac{\rho}{(\rho + \overline\xi)^2} &= (\xi - \overline\xi) \frac{\rho}{(\rho + \overline\xi'')^2} + (\xi - \overline\xi) \left\{ \frac{\rho}{(\rho + \overline\xi)^2}  - \frac{\rho}{(\rho + \overline\xi'')^2} \right\} \\
&= \left( \frac{\xi}{\rho + \xi} - \frac{\overline\xi}{\rho + \overline\xi}  \right)  + (\xi - \overline\xi) \left\{ \frac{\rho}{(\rho + \overline\xi)^2}  - \frac{\rho}{(\rho + \overline\xi'')^2} \right\} 
\end{align*}
since for $g(\xi)=\frac{\xi}{\rho + \xi}$ we have $g'(\xi) = \frac{\rho}{(\rho+\xi)^2}$.  Call the second term in the last line above $S_{2g}(x)$. 

Therefore for the first term in \eqref{eq:term1rem} we have
\begin{align*}
\int \frac{\rho}{(\rho + \overline\xi)^2}  \Big( 1 + \overline\xi\Big)^2 \Big( \mu - \overline\mu \Big) \overline\varpi \ dP_{11} &= \int \frac{\rho}{(\rho + \overline\xi)^2}  \Big( \xi - \overline\xi + S_{2f} \Big) \overline\varpi \ dP_{11} \\
&= \int \left\{ \left( \frac{\xi}{\rho + \xi} - \frac{\overline\xi}{\rho + \overline\xi} \right)  + \left( S_{2g}  +\frac{ S_{2f} \rho}{(\rho + \overline\xi)^2}   \right) \right\} \overline\varpi \ dP_{11} .
\end{align*}
The second term involving $(S_{2g}, S_{2f})$ is second-order. For the first term, note
\begin{align*}
 \int  \left( \frac{\xi}{\rho + \xi} - \frac{\overline\xi}{\rho + \overline\xi}   \right)\overline\varpi \ dP_{11} &=  \int  \left( \frac{\xi}{\rho + \xi} - \frac{\overline\xi}{\rho + \overline\xi}   \right) \Big( \overline\varpi - \varpi \Big) \ dP_{11} -  \int  \left(  \frac{\overline\xi}{\rho + \overline\xi}  - \frac{\xi}{\rho + \xi}   \right)   \ dP_{01} .
\end{align*}
Putting all of the above together gives the result.

\end{appendices}

\end{document}